\documentclass[aps,prb,reprint,superscriptaddress]{revtex4-1}

\usepackage{amsmath}
\usepackage{braket}
\usepackage{amssymb}
\usepackage{graphicx}
\usepackage{color}
\usepackage{changes}

\begin{document}

\title{Splitting of conductance resonance through a magnetic quantum dot in graphene}

\author{Nojoon Myoung}
\affiliation{Department of Physics Education, Chosun University, Gwangju 61452, Republic of Korea}
\author{Jung-Wan Ryu}
\affiliation{Center for Theoretical Physics of Complex Systems, Institute for Basic Science (IBS), Daejeon 34126, Republic of Korea}
\author{Hee Chul Park}
\affiliation{Center for Theoretical Physics of Complex Systems, Institute for Basic Science (IBS), Daejeon 34126, Republic of Korea}
\author{Seung Joo Lee}
\affiliation{Quantum-functional Semiconductor Research Center, Dongguk University, Seoul 04620, Republic of Korea}
\author{Sungjong Woo}
\email{sjwoo@ibs.re.kr}
\affiliation{Center for Theoretical Physics of Complex Systems, Institute for Basic Science (IBS), Daejeon 34126, Republic of Korea}

\begin{abstract}
We report a dual resonance feature in ballistic conductance through a quantum Hall graphene nanoribbon with a magnetic quantum dot. Such a magnetic quantum dot localizes Dirac fermions exhibiting anisotropic eigenenergy spectra with broken time-reversal symmetry. Interplay between the localized states and quantum Hall edge states is found to be two-fold, showing Breit-Wigner and Fano resonances, which is reminiscent of a double quantum dot system. By fitting the numerical results with the Fano-Breit-Wigner lineshape from the double quantum dot model, we demonstrate that the two-fold resonance is due to the valley mixing that comes from the coupling of the magnetic quantum dot with quantum Hall edge channels; an effective double quantum dot system emerges from a single magnetic quantum dot in virtue of the valley degree of freedom. 
It is further confirmed that the coupling is weaker for the Fano resonance and stronger for the Breit-Wigner resonace.
\end{abstract}

\maketitle

\section{Introduction}

Graphene has been studied extensively in past decades due to its unique electrical properties coming from the gapless and linear dispersion so-called `Dirac cone' at the corners of the 1st Brillouin zone.\cite{CastroNeto2009}  Its prominent transport behavior such as high carrier mobility\cite{Zomer2011} makes graphene promising as a candidate material to succeed silicon in the nanoelectronic industry.

Lately, graphene on a two-dimensional hexagonal boron nitride substrate has attracted exceptional interest in condensed matter physics, because it allows graphene to preserve its distinct transport properties via encapsulation.\cite{Zomer2011}  High mobility with long mean free path of such an encapsulated graphene sample enables experimental investigations on theoretical predictions which require ballistic and coherent transport over micrometer-size devices such as Veselago lens,\cite{Cheianov2007,Lee2015,Chen2016} valley-isospin dependent quantum Hall effects,\cite{Tworzydlo2007} \textit{etc.} Among other predictions, effects of inhomogeneous magnetic fields on Dirac fermion transport have been intensively and widely investigated in terms of academic interests and device applications.\cite{DeMartino2007,Kormanyos2008,Masir2008,Park2008,Ghosh2008,
Giavaras2009,Vozmediano2010,Myoung2011,Roy2012,Downing2016,
Eshghi2017,Kuru2018}

The motion of electrons at the discontinuity of external homogeneous magnetic field is confined to the
field interface.
\cite{Ghosh2008,Orszlany2008,Park2008,Taychatanapat2015} Such localization is classically
understood by snake trajectories of the motion of electrons, which corresponds to quantum Hall interface 
states at a p-n junction.\cite{Williams2007,Abanin2007,Amet2014,Klimov2015,Morikawa2015,Wei2017,
Myoung2017,Makk2018} The snake states comprise alternating semicircles with opposite chiralities, so their 
propagating direction is determined to be one-way along the discontinuous interface. 
At a 
circular boundary of  magnetic field domains, the snake trajectory of the motion of an electron can result 
in a closed orbit confining the electronic states within the circular region, a magnetic quantum 
dot~(MQD). 
Such localized states with discrete energy spectra has been known to exist within a MQD
that can be realized 
by screening homogeneous magnetic field locally.\cite{Sim1998} The snake-like localized states in a MQD mimic energy 
levels of orbitals in an atom suggesting plausible possibility of qubit architectures.

Electronic transport carrying charge or spin through quantum dots has been one of the most important topics in 
condensed matter physics, since it provides promising ways toward cutting-edge 
technology such as quantum computing
\cite{Loss1998,Steane1998,Ladd2010,Aasen2016,Godfrin2017,Watson2018} and resonant tunneling devices.
\cite{Beenakker1991,Simmel1999,Velasco2018} 
For coherent transport in quantum-dot-embedded devices, ultra-clean and low temerature evironment is 
required.
Graphene, having a long-range 2-dimensional ballistic transport property, is advantageous in fabricating coherent transport devices.\cite{Du2008,Mayorov2011,Banszerus2016,Atteia2017,Nemnes2018} Although substantial effort has been made to investigate transport through electrostatic QDs in graphene,\cite{Bardarson2009,Lee2016,Freitag2017,Wang2017,Mirzakhani2018} the research on MQDs has been relatively less explored.
It is, however, noteworthy that the lack of Klein tunneling makes the magnetic confinement is more efficient than the electrostatic confinement.\cite{DeMartino2007,Trauzettel2007,Myoung2009,Esmailpour2018}

\begin{figure}[hpbt!]
\includegraphics[width=8.5cm]{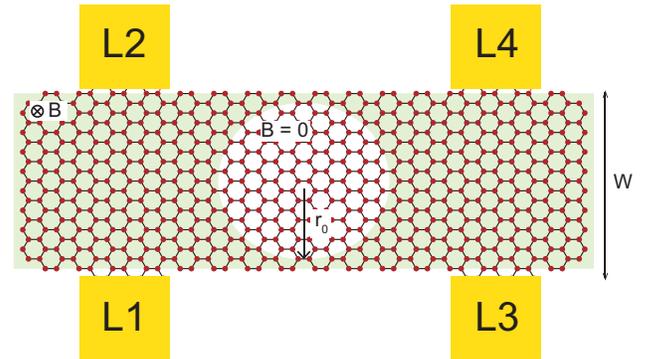}
\caption{Schematic diagram of the system considered in this work. The external magnetic field is expelled out in the circular area characterized by radius $r_{0}$. There are four leads labeled as L1, L2, L3, and L4 for scattering matrix calculations from a lead to another lead.} \label{fg:model}
\end{figure}

In the present work, we investigate electronic transport in graphene nanoribbons where a MQD is located between two quantum Hall edge channels. We analyze the feature of resonance tunnelings in the conductance through the MQD, evidencing the existence of the localized states in the MQD. The results of the analysis are discussed in the context of the coupling between the extended edge states and the discrete localized states in the MQD, showing the Breit-Wigner and Fano resonances at specific energies corresponding to the energy levels of the MQD.

The present paper is composed of the following sections. Section \ref{sec:model} outlines the MQD in a 2D graphene sheet. We offer a preliminary study of eigenenergy spectra for the MQD which indicate the formation of localized states. The continuity of the wavefunctions are assumed accross the boundary of the MQD in the analytical calculations. In Sec. \ref{sec:results}, numerical results of the ballistic conductance across the MQD in a graphene nanoribbon are analyzed. 
We compute transport properties numerically with the scattering matrix formalism based on the tight-binding method. 
Resulting resonant features in the conductance spectra are discussed addressing Fano and Breit-Wigner resonances. We additionally remark on the dot--edge-distance dependence of the energy detuning between two types of resonances. Finally, Sec. \ref{sec:conclusion} brings our work to a conclusion.

\section{Electronic properties of MQD} \label{sec:model}

\subsection{Analytic framework for model}

\begin{figure*}[hbpt!]
\includegraphics[width=17.0cm]{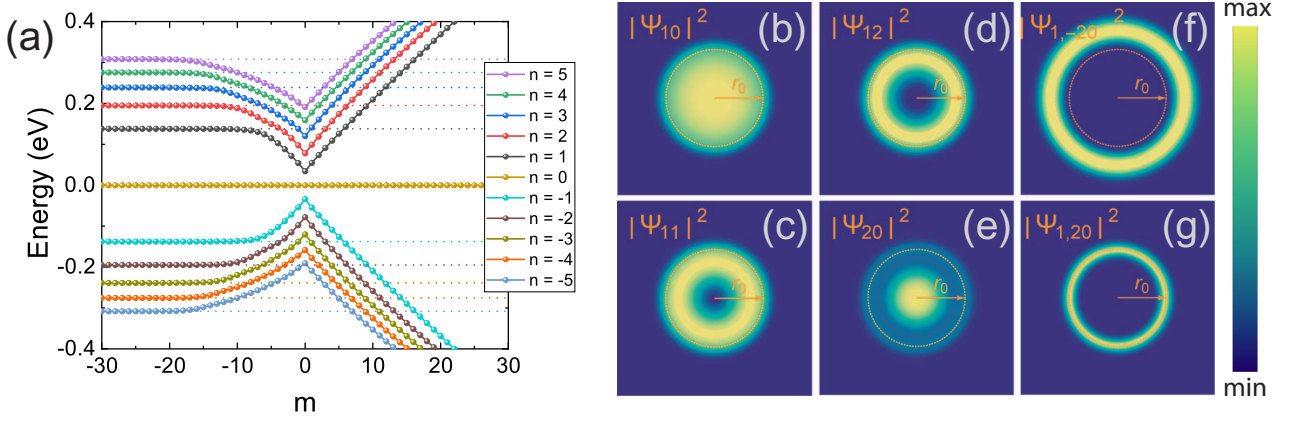}
\caption{ (a) Eigenenergy spectra $\varepsilon_{nm}$ of the magnetic quantum dot. Dotted lines indicate Landau levels of graphene under a homogneous magnetic field. Colored solid dots correspond to $\varepsilon_{nm}$ values, connected by the same-color solid lines of which colors refer different $n$ indices. (b-g) Localized state intensities of the MQD for different indices. Dotted lines indicate the boundary of the MQD.} \label{fg:eigenenergy}
\end{figure*}

We now discuss localized-state solutions in our system which is modeled as a circularly symmetric quantum dot using magnetic fields. A MQD is modeled as
\begin{align}
\vec{B}=\left\{\begin{array}{ll}B\hat{z},&r>r_{0}\\0,&r<r_{0}\end{array}\right.,
\end{align}
where $r_{0}$ is the radius of the MQD. 
Such non-uniform magnetic field can be practically realized by top-gated structures using a disk-like superconductor electrode with a thin dielectric spacer such as few-layered $h$-BN between the top local gate and graphene.

Dirac Hamiltonian of the system reads
\begin{align}
H_{\nu}&=v_{F}\left(\pi_{x}\sigma_{x}+\nu\pi_{y}\sigma_{y}\right)+U\sigma_{0},
\end{align}
where $v_{F}\simeq 10^{6}$ ms$^{-1}$ is the Fermi velocity, $\pi_i = p_i+eA_i$ is the kinetic
 momentum under magnetic field, 
$\sigma_x$ and $\sigma_y$ are Pauli matices, 
$\sigma_{0}$ is the unity matrix, and $\nu=\pm1$ for $K$ and $K'$ valleys. For simplicity, we suppose that the system is electrostatically homogeneous; $U=0$. By choosing appropriate gauge in a plane polar coordinate, the Dirac equations are written as,
\begin{align}
\label{DiracEq1}\hbar v_{F}e^{-i\phi}\left[-i\frac{\partial}{\partial r}-i\nu\left(-\frac{i}{r}\frac{\partial}{\partial \phi}+\frac{eA_{\phi}}{\hbar}\right)\right]\psi_{B,\nu}&=E\psi_{A,\nu},\\
\label{DirecEq2}\hbar v_{F}e^{i\phi}\left[-i\frac{\partial}{\partial r}+i\nu\left(-\frac{i}{r}\frac{\partial}{\partial \phi}+\frac{eA_{\phi}}{\hbar}\right)\right]\psi_{A,\nu}&=E\psi_{B,\nu},
\end{align}
where the gauge field is given by
\begin{align}
A_\phi=\left\{\begin{array}{ll}\frac{B}{2r}\left(r^{2}-r_{0}^{2}\right),&r>r_{0},\\0,&r<r_{0},\end{array}\right.
\end{align}
with $A_r = 0$ corresponding to $\vec{\nabla}\times\vec{A}=\vec{B}$. Since the system has rotational 
symmetry, the solution of the Dirac equation has the form $\Psi\left(r,\phi\right)=e^{im\phi}R(r)$,
where $m$ is an integer and $R(r)=(R_A, R_B, R_{A'}, R_{B'})^T$.\cite{Recher2009} 
In the following, all the formulas are dimensionless based on magnetic length and Landau energy gap: $l_{B}=\sqrt{\hbar/eB}$ for the length and 
$E_{0}=\sqrt{2}\hbar v_{F}/l_{B}=\sqrt{2\hbar v_{F}^{2}eB}$ for the energy.
Eliminating $\psi_{B,\nu}$ in Eqs.~(\ref{DiracEq1}) and (\ref{DirecEq2}) yields
\begin{align}
\left(\frac{d^{2}}{dr^{2}}+\frac{1}{r}\frac{d}{dr}-\frac{m^{2}}{r^{2}}+2E^{2}\right)R_{A,\nu}=0, \label{eq:indot}
\end{align}
for $r<r_{0}$ and
\begin{align}
&\left(\frac{d^{2}}{dr^{2}}+\frac{1}{r}\frac{d}{dr}-\frac{m_{eff}^{2}}{r^{2}}-\frac{1}{4}r^{2}-2 m_{eff}-2\nu+2E^{2}\right)\nonumber\\
&\times R_{A,\nu}=0, \label{eq:outdot}
\end{align}
for $r>r_{0}$, where $m_{eff}\equiv m-s$. Here, $s\equiv B\pi r^{2}_{0}e/h$ is the 
`missing' flux, which indicates the amount of magnetic flux 
screened out from the MQD. For $r<r_{0}$, the solution of the differential equation is the Bessel's function of the first kind,
\begin{align}
\psi_{A,\nu}\left(r,\phi\right)=c_{1}J_{\left|m\right|}\left(\sqrt{2}E r\right)e^{im\phi},
\end{align}
while the solution for $r>r_{0}$ is 
\begin{align}
\psi_{A,\nu}\left(r,\phi\right)=c_{2}r^{\left|m_{eff}\right|}e^{-r^{2}/4}U\left(a,b;r^{2}/2\right)e^{im\phi}.
\end{align}
Here, $U\left(a,b;r^{2}/2\right)$ is the confluent hypergeometric function with
\begin{align}
a&=\frac{\left|m_{eff}\right|+m_{eff}+\left(\nu+1\right)}{2}-E^{2},\\
b&=\left|m_{eff}\right|+1.
\end{align}
Normalized coefficients $c_{1}$ and $c_{2}$ are determined from the continuity condition. 
Furthermore, $\psi_{B,\nu}$ can be obtained using Eq.~(\ref{DirecEq2})
\begin{align}
\psi_{B,\nu}\left(r,\phi\right)=i\nu c_{1}J_{\left|m\right|+\nu}\left(\sqrt{2}E r\right)e^{i\left(m+1\right)\phi},
\end{align}
for $r<r_{0}$, and
\begin{align}
&\psi_{B,\nu}\left(r,\phi\right)=i\frac{c_{2}}{E}r^{\left|m_{eff}\right|-1}e^{-r^{2}/4}e^{i\left(m+1\right)\phi}\nonumber\\
&\qquad\times\left\{\left(-\left|m_{eff}\right|+\nu m_{eff}+\frac{\nu+1}{2}r^{2}\right)\right.U\left(a,b;r^{2}/2\right)\nonumber\\
&\qquad\left.+ar^{2}U\left(a+1,b+\nu;r^{2}/2\right)\right\},
\end{align}
for $r>r_{0}$.

\subsection{Eigenenergies and localized state wavefunctions}

The continuity of $R(r)$ at $r=r_0$ gives
\begin{align}
\left[\begin{array}{cc}m_{11}&m_{12}\\m_{21}&m_{22}\end{array}\right]\left[\begin{array}{cc}c_{1}\\c_{2}\end{array}\right]=0,
\end{align}
where
\begin{align}
&m_{11}=J_{\left|m\right|}\left(\sqrt{2}E r_{0}\right),\qquad m_{21}=i\nu J_{\left|m\right|+\nu}\left(\sqrt{2}E r_{0}\right)\nonumber\\
&m_{12}=-r_{0}^{\left|m_{eff}\right|}e^{-r_{0}^{2}/4}U\left(a,b;r_{0}^{2}/2\right)\nonumber\\
&m_{22}=-\frac{i}{E}r_{0}^{\left|m_{eff}\right|-1}e^{-r_{0}^{2}/2}\nonumber\\
&\times\left\{\left[-\left|m_{eff}\right|+\nu m_{eff}+\frac{1+\nu}{2}r_{0}^{2}\right]U\left(a,b;r_{0}^{2}/2\right)\right.\nonumber\\
&\left.-ar_{0}^{2}U\left(a+1,b+\nu;r_{0}^{2}/2\right)\right\}.
\end{align}
For non-trivial solutions, we numerically solve the secular equation, $m_{11}m_{22}-m_{12}m_{21}=0$, and find eigenenergies $E_{nm}$ with integers $n$ and $m$ for each real value of $s$. The resulting eigenenergies are plotted in Fig.\ref{fg:eigenenergy}. It is noted that the two valleys are degenerate in the eigenenergy spectra in case of an isolated MQD in an infinite system.

Due to the broken time-reversal symmetry, the eigenenergies exhibit asymmetric behavior with respect to the sign of $m$. 
For small $|m|$ there are small discrepancies between time-reversal partners, $E_{nm}$ and $E_{n,-m}$. 
It is due to the fact that the wavefunctions with small $|m|$ are mainly resides within the quantum dot where the magnetic field is zero so that the time-reversal symmetry is barely broken. 
On the other hand, for larger $|m|$, larger discrepancies are observed between time-reversal partners.

One noticeable analysis is the direction of the persistent current along the MQD boundary, defined by $I_{nm}=\left(1/\hbar\right)\partial E_{nm}/\partial m$. The signs of $I_{nm}$ indicates that the semiclassical trajectories for the localized states can be either clockwise or counter-clockwise rotation of snake orbits along the MQD boundary.\cite{Sim1998}. Interestingly, there is a flat eigenenergy spectrum at zero energy, not depending on $m$. The zero-energy states are understood by solving the differential equation with $E=0$. The resulting eigenstates of the zero energy are given by a linear summation of electron states on $A(B)$ sites and hole states on $B(A)$ sites.\cite{Grujic2011}

Analysis of each energy eigenstate $\Psi_{nm}$ corresponding to $E_{nm}$ shows that $n$ and $m$ are the radial and angular
quantum numbers, respectively.
Each eigenstate $\Psi_{nm}$ describes how the localized states are formed in the MQD. 
Some of the localized state wavefunctions are plotted in Fig. \ref{fg:eigenenergy}(b-g). Note that the intensities of the 
wavefunctions are isotropic but dependent on quantum numbers $n$ and $m$. The wavefunctions for $m=0$ ($
\Psi_{10}$ and $\Psi_{20}$) exhibit their maxima at the center of the MQD, whereas the wavefunctions for $m\neq0$ 
($\Psi_{11}$ and $\Psi_{12}$) show their minima at the MQD center. Especially, as displayed in Fig. 
\ref{fg:eigenenergy}(f), the localized state wavefunctions for $m=-20$ are found totally outside the MQD boundary, 
forming cyclotron orbits in a uniform-magnetic-field region. 
As we already mentioned, this is consistent with the finding 
that $E_{nm}$ converges to $n$-th Landau level as $|m|$ increases with $m<0$. 
The slope of the linear dispersion can be estimated to be the kinetic energy of a Dirac
particle located near the boundary of the QD, $E\sim m/(\sqrt{2}r_0)$
The critical value for $m$ where the dispersion changes from linear to flat Landau levels can be estimated by equating the linear dispersion with the corresponding Landau level energy which is identity in our magnetic unit for $n=1$ so that $m_c\sim\sqrt{2}r_0\approx 10$.
This value matches reasonably well with the dispersions in Fig. \ref{fg:eigenenergy}(a).
On the other hand, in Fig. \ref{fg:eigenenergy}(g), $\left|\Psi_{1,20}\right|^{2}$ exhibits distinct feature to $
\Psi_{1,-20}$ because of the broken time-reversal symmetry. 
The localized state wavefunction 
for $m=20$ mainly reside in the MQD.

\section{Ballistic Conductance through Magnetic Quantum Dot} \label{sec:results}

So far, we have studied the eigenenergies and localized state wavefunction of the isolated MQD in an infinitely large graphene sheet. In practice, as depicted in Fig. \ref{fg:model}, the sizes of graphene samples for conductance measurements are finite, and quantum Hall edge channels are formed along the edge of the sample.
For a MQD on such a finite-sized sample, the localized states are inevitable to have coupling with 
the extended edge states of the sample, leaving resonances in the conductance spectrum.
In this study, we use a 98.4 nm-wide graphene nanoribbon with armchair edges to take into account the valley-isospin-dependent quantum Hall effects.\cite{Tworzydlo2007} 
A magnetic quantum dot is introduced at the center of the nanoribbon with a radius of 44.3 nm which is about 5 nm apart from both edges. 
In our simulations, we have used $B=$ 15.7 T for which the cyclotron radius $r_{c}=$ 6.5 nm, so that the aformentioned coupling 
is substantial. 
As a consequence of the coupling, Dirac fermions coming from one edge can be transferred to the other edge through the MQD.  
The conductance from L1 to L2 in Fig. \ref{fg:model}  is calculated using \textsc{kwant}
package\cite{Amestoy2001,Groth2014} in order to check 
how Dirac fermions from one edge to the other pass through the MQD. 

\begin{figure}
\includegraphics[width=8.5cm]{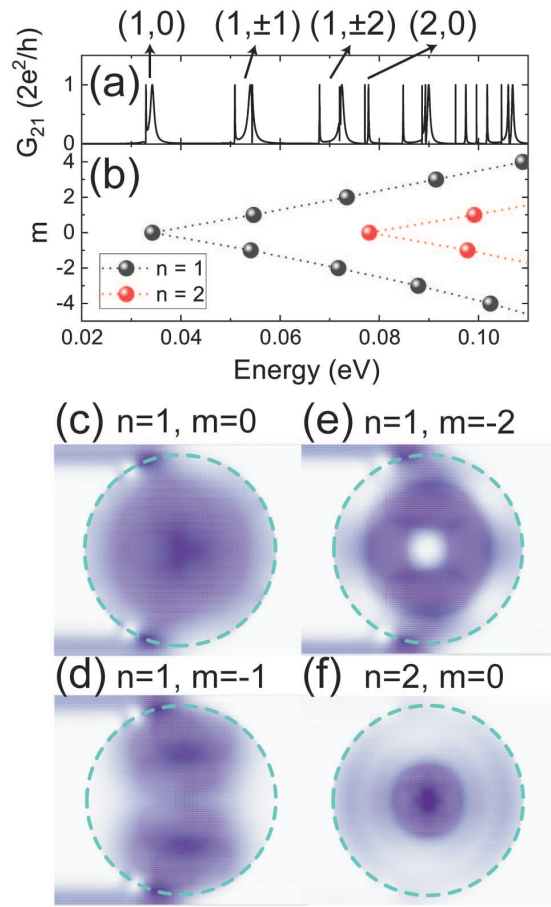}
\caption{ (a) Conductance calculated between L1 and L2 leads as a function of Dirac fermion energy. First four resonances are denoted by their corresponding localized states labelled as $\left(n,m\right)$. (b) Enlargement of the eigenenergy spectra focussing on the lower eigenenergies. Black and red symbols represent the eigenenergy bands for $n=1$ and $n=2$, respectively. Parameter $r_{0}=4.2$ is used in the eigenenergy calculation. (c-f) Image plots of wavefunctions of the scattering region in $S$-matrix formalism acquired from \textsc{KWANT} codes, for given Dirac fermion energies corresponding to the first four resonances in the conductance spectra (a). The dashed line indicates the size of the MQD. Note that the image plots are normalized by their maximum values for clarity of viewing.}\label{fg:cond}
\end{figure}

\begin{figure*}
\includegraphics[width=14cm]{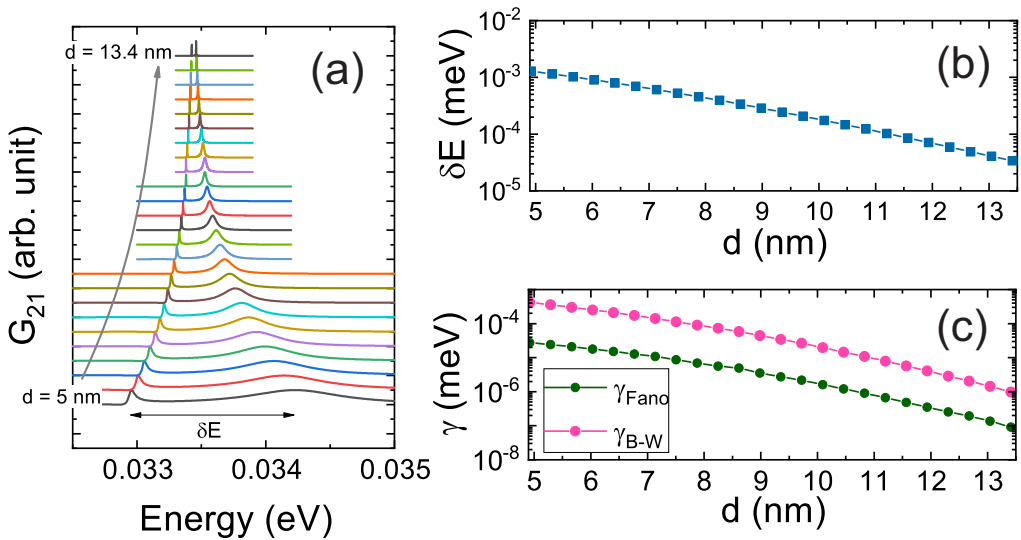}
\caption{ (a) Conductance spectra around the first resonance peaks as functions of Dirac fermion energy for various edge-dot distances from $d=5$ nm to 13.4 nm with an increment 0.37 nm. (b) The energy splitting $\delta E$ as a function of $d$ from the fitted data. (c) Coupling strengths of Fano and Breit-Wigner resonances, $\gamma_{\rm Fano}$ and $\gamma_{\rm BW}$, from the numerical fitting as functions of $d$.} \label{fg:detuning}
\end{figure*}

Figure \ref{fg:cond} presents the conductance as a function of 
the energy of Dirac fermions. 
One can see that there are a number of resonance peaks. 
Without MQD the conductance should be zero since two quantum Hall edge channels cannot talk to each other in the ballistic regime. Figure \ref{fg:cond}(a) and (b) show that energies of the resonance peaks match well to the eigenenergies of the localized states in the MQD. Such correspondence implies that the conductance resonances are indeed consequences of the coupling between the edge channels and the localized states on the MQD. 
Figure \ref{fg:cond}(c)-(f) demonstrate that each resonance does result from the localized states in the MQD with a good agreement with the analytic solutions given in Fig. \ref{fg:eigenenergy}(b)-(e). 

Detailed look into Fig.~\ref{fg:cond} further shows  
interesting features 
within the resonance peaks.
There are two distinct shapes of resonances for each localized states,
broad and symmetric peaks for the 
Breit-Wigner resonances and sharp and asymmetric peaks for the Fano resonances. 
A Breit-Wigner resonance occurs when an extended state
is strongly coupled with localized states.
On the other hand, a Fano resonance 
is an interference effect due to the weak coupling
between extended and localized states. 
Such conductance spectra are discovered in charge transport through coupled double quantum 
dots, where Breit-Wigner and Fano resonances occur at energies corresponding to
bonding and antibodnding of the two dots.\cite{Guevara2003}
In this study, even though there is only one MQD, there are two degenerate valley states, which are split with valley mixing due to the coupling with the edge channels
\cite{Akhmerov2007,Tworzydlo2007,Sekera2017,Handschin2017}. 
Quantitative analysis of the coexistence of Fano and Breit-Wigner resonances is done
by performing numerical fitting using the double QD model,
\cite{Guevara2003} finding the peak positions, $E_{\rm Fano}$ and $E_{\rm BW}$, and their widths,
$\gamma_{\rm Fano}$ and $\gamma_{\rm BW}$.

We first note that the Fano peak in a peak splitting on a MQD has lower energy 
compared to the Breit-Wigner peak, which is contrary to a general double QD.
In a double QD, Fano peak is for the antibonding state of two dots which has higher energy than 
the bonding.
For a graphene nonoribbon with armchair edges allows only antisymmetric states of valleys, 
$\psi_K-\psi_{K'}$.\cite{Sasaki2011, Brey2006}
With a MQD, although symmetric states of two valleys are allowed 
due to the partial existence of zigzag edge along the boundary of the dot, 
antisymmetric states are still preferred energetically, i.e., have lower energy.
Symmetric valley mixing has disbenefit of kinetic energy at the armchair edge while antibonding
of two dots in the double QD model has disbenefit of kinetic energy at the node of wavefunction between the two dots.

In order to further understand the valley splitting in the MQD with the edge channels, we investigate how the conductance spectra for the resonances behave depending on the distance between the MQD and the edges, $d$. Figure \ref{fg:detuning}(a) shows the $d$-dependence of the conductance resonances in case of $\left(n,m\right)=\left(1,0\right)$. 
It is clearly seen that the Fano and Breit-Wigner resonant peaks become closer to each other as $d$ increases, eventually converging to a single peak for sufficiently large $d$. 
It is because the MQD gets decoupled from the edge.
The exponential behaviors of the resonant peak splitting, $\delta E = E_{\rm BW} - E_{\rm Fano}$, and coupling strengths, $\gamma_{\rm Fano}$ and $\gamma_{\rm BW}$, in terms of $d$ [Fig. \ref{fg:detuning}(b) and (c)] imply
that such valley splitting stems from the coupling with the edge;
the coupling strength between the MQD and the edge is proportional to wavefunction overlap between the localized states and the edge channels. 
Figure \ref{fg:detuning}(c) also confirms the fact that the coupling for the Fano resonance should be much weaker than that for the Breit-Wigner resonances.
The two-level splitting for a single MQD is a unique feature of graphene nanoribbon with the presence of valley isospin degeneracy.

\section{Conclusions} \label{sec:conclusion}

In conclusion, using both numerical and analytical approaches, we have shown that the quantum Hall conductance
on a graphene sample with a MQD exhibits two distinguished resonant spectra, Fano and Briet-Wigner resonances, as a consequence of the valley mixing in the MQD. 
Even though an isolated MQD has valley degeneracy, the coupling between the MQD and edge channels leads to valley mixing for a finite-size quantum Hall graphene system. 
By fitting the resonant spectra, we have demonstrated that the two-level splitting due to the valley mixing becomes smaller as the distance between the MQD and edge channels increases, accompanied with narrower spectral widths of resonances. It shows that the valley mixing is due to the wavefunction overlap between the MQD and the edge channel. 
Analysis of the numerical results has confirmed that the coupling for the Fano resonances is much weaker than that for the Breit-Wigner resonances. 
When compared to conventional two-dimensional electron gas systems, the coexistence of Fano and Breit-Wigner resonances for a single dot structure on a graphene nanoribbon is a unique phenomena, which is due to the valley degree of freedom. 
The reason why the Fano peak in a MQD has lower energy 
than the Breit-Wigner peak differently from a double QD model is also explained from the perspective of edge coupling.
The two-level MQD system with a strong coupling with the edge channels may inaugurate both theoretical and experimental research on non-Hermitian electronic systems.
Moreover, our findings regarding the conductance resonances possess potential applications such as graphene-based gas or chemical sensors, utilizing possible high sensitivity with the sharp resonances.

\acknowledgments

This work is supported by the National Research Foundation of Korea (NRF) grant and Korea Institute for Advanced Study(KIAS) funded by the Korea government (MSIT and MOE) (2017R1C1B5076824, Project IBS-R024-D1,NRF2016-R1D1A1B04-935798), and Chosun University (2017).

\bibliography{MQDGra}

\end{document}